\documentclass[conference]{IEEEtran}

\IEEEoverridecommandlockouts

\usepackage[utf8]{inputenc}
\usepackage[T1]{fontenc}
\usepackage[nopatch=footnote]{microtype}
\usepackage{textcomp}
\usepackage{amsmath}
\usepackage{amsfonts}
\usepackage{bbm}
\usepackage{algorithmic}
\usepackage{algorithm}
\usepackage{graphicx}
\usepackage{textcomp}
\usepackage{xcolor}
\usepackage{xspace}
\usepackage{hyperref}
\usepackage{cleveref}
\usepackage{setspace}
\usepackage{wrapfig}
\usepackage{multirow}
\usepackage{balance}
\usepackage{rotating}
\usepackage[font=footnotesize]{caption}
\usepackage{hhline}

\balance

\def\tablename{Table}
\renewcommand{\thetable}{\arabic{table}}

\def\BibTeX{{\rm B\kern-.05em{\sc i\kern-.025em b}\kern-.08em
    T\kern-.1667em\lower.7ex\hbox{E}\kern-.125emX}}
\begin{document}

\title{Using Fourier Analysis and Mutant Clustering to Accelerate DNN Mutation Testing}

\author{\IEEEauthorblockN{Ali Ghanbari}
\IEEEauthorblockA{
\textit{Auburn University}\\
Auburn, AL, USA\\
\url{ghanbari@auburn.edu}
}

\and

\IEEEauthorblockN{Sasan Tavakkol}
\IEEEauthorblockA{
\textit{Google Research}\\
Irvine, CA, USA\\
\url{tavakkol@google.com}
}
}

\crefformat{section}{\S#2#1#3}
\crefformat{subsection}{\S#2#1#3}
\crefformat{subsubsection}{\S#2#1#3}

\newcommand{\totalDNNs}{14\xspace}

\newcommand{\dms}{DM\#\xspace}
\newcommand{\noFFT}{$\text{DM\#}_\star$\xspace}

\newcommand{\ie}{\textit{i.e.}\xspace}
\newcommand{\eg}{\textit{e.g.}\xspace}
\newcommand{\etc}{\textit{etc}\xspace}
\newcommand{\perse}{\textit{per se}\xspace}
\newcommand{\ala}{\textit{à la}\xspace}
\newcommand{\cf}{\textit{c.f.}\xspace}
\newcommand{\via}{\textit{via}\xspace}
\newcommand{\vs}{\textit{vs.}\xspace}
\newcommand{\etal}{\textit{et al.}\xspace}
\newcommand{\viceversa}{\textit{vice versa}\xspace}

\newcommand{\ali}[1]{\textcolor[rgb]{0.0,0.0,1.0}{#1}}
\newcommand{\sasan}[1]{\textcolor[rgb]{1.0,0.0,0.0}{#1}}

\renewcommand{\algorithmicrequire}{\textbf{Input:}}
\renewcommand{\algorithmicensure}{\textbf{Output:}}

\newtheorem{definition}{Definition}

\maketitle

\thispagestyle{plain}
\pagestyle{plain}

\begin{abstract}
Deep neural network (DNN) mutation analysis is a promising approach to evaluating test set adequacy.
Due to the large number of generated mutants that must be tested on large datasets, mutation analysis is costly.
In this paper, we present a technique, named \dms, for accelerating DNN mutation testing using Fourier analysis.
The key insight is that DNN outputs are real-valued functions suitable for Fourier analysis that can be leveraged to quantify mutant behavior using only a few data points.
\dms uses the quantified mutant behavior to cluster the mutants so that the ones with similar behavior fall into the same group.
A representative from each group is then selected for testing, and the result of the test, \eg, whether the mutant is killed or survived, is reused for all other mutants represented by the selected mutant, obviating the need for testing other mutants.
\totalDNNs DNN models of sizes ranging from thousands to millions of parameters, trained on different datasets, are used to evaluate \dms and compare it to several baseline techniques.
Our results provide empirical evidence on the effectiveness of \dms in accelerating mutation testing by 28.38\%, on average, at the average cost of only 0.72\% error in mutation score.
Moreover, on average, \dms incurs 11.78, 15.16, and 114.36 times less mutation score error compared to random mutant selection, boundary sample selection, and random sample selection techniques, respectively, while generally offering comparable speed-up.
\end{abstract}


\section{Introduction}\label{sec:intro}
Deep learning~\cite{bib:lecun2015deep}, enabled by \textit{deep neural networks} (DNNs), has been used in modern software systems in many domains. 
The growing applications of DNNs in safety and mission critical systems, such as autonomous driving, healthcare, and energy, and the need for accuracy and robustness of such systems, warrant research on quality assurance of DNNs.
Among many methods for quality assurance of DNNs~\cite{bib:liu2021algorithms,bib:huang2020survey}, testing is a widely used approach~\cite{bib:zhang2020machine}, wherein test data points are manually curated, or automatically generated, to satisfy certain test requirements.
However, good performance on the test dataset does not necessarily imply the robustness and generality of a DNN model, and a systematic way for assessing the quality of the test data is needed.

In recent years, \textit{mutation analysis}~\cite{bib:demillo1978hints,bib:hamlet1977testing} has been reintroduced in the context of DNNs~\cite{bib:shen2018munn,bib:ma2018deepmutation,bib:humbatova2021deepcrime,bib:lu2022towards} as a promising method for assessing the quality of test data (see~\cref{sec:back:mutation} more details).
Despite its promise for test data quality assurance, DNN mutation analysis remains prohibitively expensive due to its need for testing many mutants on large datasets, limiting its practical deployment~\cite{bib:ma2018deepmutation,bib:feng2022mutation,bib:wang2023fine,bib:ghanbari2023mutation,bib:li2022higher,bib:shen2021boundary,bib:ghanbari2024decomposition}.
In the realm of conventional programs, \ie, programs that are not strictly based on data-driven pipelines, there is a large body of work concerning acceleration of mutation analysis~\cite{bib:pizzoleto2019systematic,bib:usaola2010mutation}.
However, since mutation analysis is relatively new for DNNs, this important topic, which could benefit many useful techniques relying on mutation analysis, has not received as much attention.

We present a novel method, named \dms, for accelerating DNN mutation analysis by \textit{testing fewer mutants}, which can be applied to both model-level and source-level mutation analysis.
This is achieved through efficiently clustering the mutants into sets of behaviorally similar mutants and testing representatives of each cluster to approximate the mutation score.
Our key insight is that DNN outputs, being real-valued and continuous, lend themselves naturally to Fourier analysis~\cite{bib:sneddon1995fourier,bib:tolstov2012fourier}.
This allows us to characterize the behavior of each mutant with only a handful of test samples.
Specifically, we use Fourier analysis to compute a comparable signature of the overall behavior of the mutants using a \textit{tiny fraction} of the test dataset.
This way, \dms reasons about the behavior of the mutants, \eg, whether they are likely killed or survived, solely based on their Fourier analysis results, obviating the need for the more expensive process of applying the mutants on the entire test dataset.

Given a set of mutants to be tested, \dms uses a tiny subset of the test dataset (\eg, 0.1\% of the data points) to calculate discrete Fourier transform spectra \via an efficient algorithm, known as fast Fourier transform~\cite{bib:oppenheim2009dsp} (FFT).
FFT spectra uniquely encode the behavior of functions as vectors of frequency bins and their amplitudes. 
\dms uses the similarity between FFT spectra as a metric for quantifying behavioral similarity of the corresponding mutants, which is then used to cluster the mutants into sets with likely identical mutation results, \eg, all killed or all survived.
Finally, a representative mutant from each cluster is selected for full testing, \ie, testing with all the test data points.
The mutation testing result of the representative is then reused for all the mutants it represents. 

We have implemented \dms~\cite{bib:replica}, which is applicable to a wide range of supervised classifier DNN architectures, such as fully-connected neural networks (FCNNs), convolutional neural networks (CNNs) with/without residual blocks, and recurrent neural networks (RNNs).
\dms complements the existing work~\cite{bib:feng2022mutation,bib:wang2023fine,bib:ghanbari2023mutation,bib:li2022higher,bib:shen2021boundary,bib:ghanbari2024decomposition} and can be used alongside them for further acceleration.
To increase its usability as a tool, we have designed \dms not to require any parameters from the user to operate, instead it leverages lightweight search procedures for finding appropriate parameter values for FFT analysis, as well as clustering (see~\cref{sec:approach} for more details).

We empirically study \dms using a set of \totalDNNs DNN models of varying sizes, ranging from thousands to millions of parameters, representative of a broad range of models with FCNN, CNN, and RNN architectures. 
Our experiment with \dms consists of four parts.
First, we use a small subset of our benchmark models, consisting of simple FCNN models, to provide empirical evidence that \dms components are well-behaved.
As a byproduct of this phase, we also obtain heuristic default values for guiding the parameter search procedures such that \dms incurs a maximum of 5\% error in mutation score and achieves a minimum of 10\% speed-up.

Next, we apply \dms, with its default parameters, to another subset of benchmark models containing larger and more complex DNN models, representative of real-world DNN models.
The evaluation results provide empirical evidence on the effectiveness of \dms in reducing mutation testing time without incurring significant error in mutation score.
Specifically, \dms reduces the number of mutants to be tested by 35.71\%, on average, which translates into 28.38\% reduction in end-to-end mutation testing time (when the overhead of \dms itself is also taken into account), while incurring only an average of 0.72\% error in mutation score.
Furthermore, we compare \dms to three baselines techniques: random mutant selection~\cite{bib:ghanbari2023mutation} (RMS), boundary sample selection~\cite{bib:shen2021boundary} (BSS), and random sample selection (RSS).
We observed that while RMS, BSS, and RSS are 1.28, 2.38, 2.91 times, respectively, faster than \dms, \dms incurs 11.78, 15.16, and 114.36 times less mutation score error than RMS, BSS, and RSS, respectively.

Lastly, we examine whether FFT-based clustering in \dms captures meaningful behavioral similarities rather than reducing to mutant output histograms, and whether propagating test outcomes from cluster representatives to other mutants remains reliable.
Skipping FFT analysis increases mutation score error by over $24\times$.
We also confirm that clustering results can be reliably propagated from representatives to other mutants (see~\cref{sec:experiments} for details).

In summary, this paper makes the following contributions.
\begin{itemize}
    \item \textbf{Concept:} We introduce the concept of applying FFT to quantify the behavior of DNN mutants with few data. While we have explored clustering of the mutants induced by FFT spectra for accelerating mutation testing, many other applications, \eg, model compression and inference optimization, are also conceivable.
    \item \textbf{Implementation:} We have implemented \dms~\cite{bib:replica}, which can be used as a mutation testing add-on for the existing frameworks, such as DeepMutation~\cite{bib:ma2018deepmutation} or DeepCrime~\cite{bib:humbatova2021deepcrime}, with a bit of integration coding.
    \item \textbf{Empirical evaluation}: We present an empirical evaluation of \dms on a diverse set of DNN models, and a comparison to the related work. \dms reduces mutation testing time with negligible error in mutation score. It also outperforms baseline techniques in terms of mutation score accuracy, while offering comparable speed-up.
\end{itemize}

\section{Background}\label{sec:background}
\subsection{Mutation Analysis}\label{sec:back:mutation}
Mutation analysis~\cite{bib:demillo1978hints,bib:hamlet1977testing} is a program analysis method for assessing test suite quality in conventional programs.
Central to this method is the set of program transformation operators known as \textit{mutation operators}, or \textit{mutators}.
Mutation analysis involves generating a set of program variants, called \textit{mutants}, by systematically mutating program elements using mutators, and running the test suite on the mutants to check if the outputs of the mutants are different from that of the original program; if different, the mutant is marked as \textit{killed}, otherwise as \textit{survived}.
Test suite quality is measured by \textit{mutation score}, which is traditionally approximated by the ratio of killed mutants over survived mutants.
Higher mutation scores indicate stronger suites.
While mutation analysis, or its acceleration, does not directly find bugs, it helps strengthening testing by measuring and improving test suite quality.
Beyond its original application in assessing the quality of test suites, mutation analysis has had many other applications~\cite{bib:papadakis2019mutation,bib:jia2010analysis}.

Recently mutation analysis has been applied in the context of DNNs~\cite{bib:shen2018munn,bib:ma2018deepmutation,bib:humbatova2021deepcrime,bib:lu2022towards}.
Like its conventional counterpart, since its introduction as a test data quality assessment method, DNN mutation analysis has had plenty of applications, including adversarial sample detection~\cite{bib:wang2019adversarial} and generation~\cite{bib:hu2023muten}, robustness analysis~\cite{bib:hu2019deepmutation++,bib:lin2022robustness}, aiding manual labeling of test data \via prioritization of test data~\cite{bib:wang2021prioritizing}, accuracy estimation to alleviate the need for manual labeling of test data~\cite{bib:hu2023aries}, fault localization~\cite{bib:ghanbari2023mutation}, automated repair of DNNs~\cite{bib:sohn2023arachne,bib:wu2022genmunn}, modular decomposition of DNN models~\cite{bib:ghanbari2024decomposition}, and improving the quality of test dataset by generating new data points guided by mutation testing~\cite{bib:riccio2021deepmetis,bib:deokuliar2023improving,bib:zohdinasab2024focused}.
%

DNN mutation analysis is usually done at two different levels: (1) source code and the data used to specify and train a model, known as \textit{source-level mutation analysis}; (2) the graph representing the trained model itself, known as \textit{model-level mutation analysis}.
Both forms of mutation analysis are costly~\cite{bib:hu2019deepmutation++,bib:jahangirova2020empirical}, so there is a recent research trend in accelerating this process~\cite{bib:feng2022mutation,bib:wang2023fine,bib:ghanbari2023mutation,bib:li2022higher,bib:shen2021boundary,bib:ghanbari2024decomposition,bib:lyons2025on}.
The two forms differ from each other mainly in the way the mutants are generated, the former involves mutating the source and/or the training data and training the resulting mutants, while the latter directly mutates an already trained model.
The testing of the mutants in both forms are identical to each other, so an approach for reducing the costs of mutation testing is readily applicable in both contexts.

To study the impact of the presented acceleration technique on the mutation score, we use the mutation score formula by Ma \etal~\cite{bib:ma2018deepmutation}: $\frac{1}{|M|\times|L|}\sum_{\mu\in M}|\mbox{killingLabels}(\mu)|$, where $M$ is the set of mutants that is obtained by applying a set of mutators on a given DNN model $m$.
$L=\{\mbox{label(t)}\mid t\in T\}$ is the set of labels in a test dataset $T$, wherein the function `$\mbox{label}$' returns the ground-truth label for the data points in $T$.
For any mutant $\mu\in M$, $\mbox{killingLabels}(\mu)$ is defined to be $\{\mbox{label}(t)\mid t\in T~\mbox{and}~\mbox{kill}(\mu, t)\}$.
A mutant $\mu$ is said to be \textit{killed} by a test data point $t\in T$, denoted by the predicate $\mbox{kill}(\mu,t)$, if $\mbox{argmax}(m(t))=\mbox{label}(t)$ and $\mbox{argmax}(\mu(t))\neq\mbox{label}(t)$, where $\mbox{argmax}(m(t))$, or $\mbox{argmax}(\mu(t))$, denotes the label predicted by the model $m$, or its mutant $\mu$, for $t$.

\subsection{Fourier Analysis}\label{sec:back:fourier}
\textit{Fourier series} represent periodic, piece-wise continuous functions using a series consisting of a constant term and infinitely many sine and cosine terms.
The sine and cosine terms are harmonically related, \ie, their frequencies are integer multiples, or harmonics, of the so-called fundamental frequency, which is calculated based on the period of the original function.
The \textit{Fourier transform} extends this frequency-domain representation to piece-wise continuous functions that are not periodic.
In short, Fourier transform transforms a given piece-wise continuous function $f(\bar{x})$ into a complex-valued continuous function $\hat{f}(\omega)$ over frequencies.
This new function represents the amplitudes and phases of the different frequency components $\omega$ that make up the original function $f(\bar{x})$.
\textit{Discrete Fourier transform} is the discrete counterpart of Fourier transform that is used to calculate discrete frequency components based on a set of values sampled from the original function $f(\bar{x})$.
This notion further extends frequency-domain representation to discrete and non-continuous functions.
This paper uses \textit{fast Fourier transform} (FFT), which is an efficient (and famously elegant) implementation of discrete Fourier transform.
The output of the FFT is often called \textit{spectra}, as it defines the magnitude of each frequency bucket.
The original function can be uniquely reconstructed from the FFT spectra through a process known as \textit{inverse FFT}.
Readers are referred to~\cite{bib:tolstov2012fourier,bib:sneddon1995fourier,bib:oppenheim2009dsp} for more information about Fourier analysis.

FFT has extensively been used in understanding wave-forms, solving differential equations, image compression, audio processing, and even DNN compression and optimization~\cite{bib:raghu2017svcca,bib:lin2018fftbased,bib:lavin2016fast}.
This paper, for the first time, proposes the use of FFT in accelerating mutation analysis of DNNs.
A key motivation behind using FFT, rather than other methods~\cite{bib:kullback1951information,bib:lin1991divergence,bib:klabunde2023similarity}, is that FFT converges very quickly and enables representing a holistic view of the behavior of a complex function with only a handful of samples.

\section{Motivating Example}\label{sec:motivation}
\begin{figure}[t]
    \centering
    \resizebox{\columnwidth}{!}{
    \includegraphics{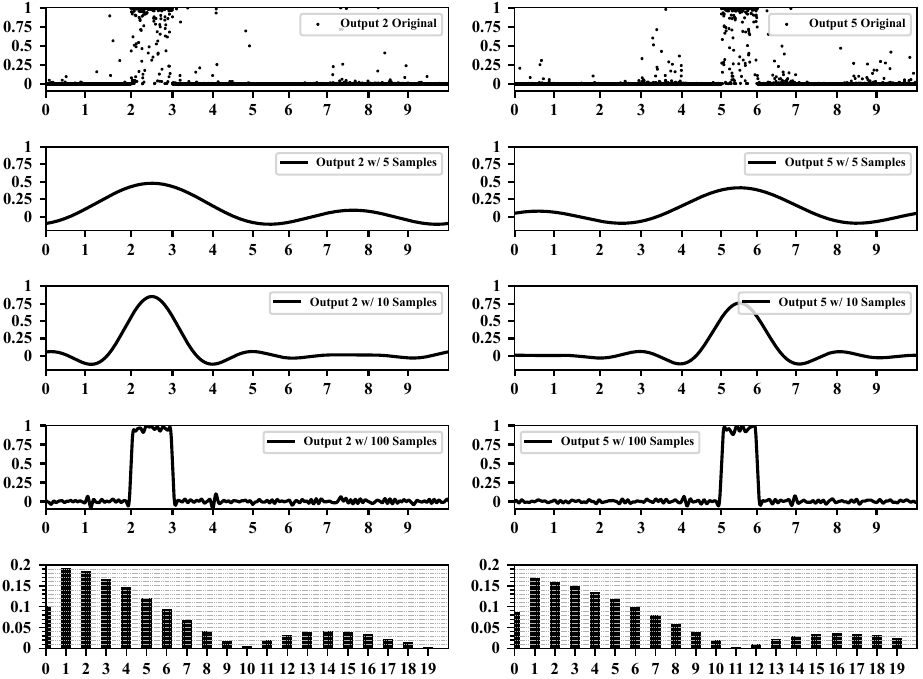}
    }
    \begin{scriptsize}
        \begin{tabular}{p{4cm}p{4cm}}
            \qquad\qquad\qquad\qquad(a) & \qquad\qquad\qquad\qquad(b)
        \end{tabular}
    \end{scriptsize}
    \caption{Row 1: outputs 2 and 5 of a DNN classifier trained on MNIST dataset in columns (a) and (b), respectively. Rows 2-4: Fourier approximation of the two functions using 5, 10, and 100 terms, respectively. Last row: bar charts for the first 20 frequency buckets of the FFT for the two functions. The bar charts are annotated with gray guidelines to aid visual comparison of the heights of the bars between two diagrams.}
    \label{fig:motivating-plots}
\end{figure}

The main idea behind \dms is that FFT spectra can be used to efficiently quantify the behavior of complex real-valued functions such as DNNs.
This enables comparing different DNNs, \eg, different mutants of a DNN, to identify the ones that behave similarly, \eg, they are likely to survive or be killed by certain inputs.
To understand this idea better, let us look into a simplified example.
Assume that we have trained an FCNN model with four layers, each with 50 neurons with ReLU activation function.
After training the model using MNIST dataset, it achieves a test accuracy of 0.9727.
This model has 10 softmax outputs; the first row in Fig.~\ref{fig:motivating-plots} plots the third and sixth outputs of the model (corresponding to the output/class indices 2 and 5, respectively) in a plane where the x-axis comprises 10,000 MNIST test images arranged and sorted in ascending order of their labels, and the y-axis ranges between 0 and 1, representing softmax values.

Assume further that we take FFT of these two outputs using 5, 10, and 100 randomly selected samples that include at least one data point of the class corresponding to the plotted output.
The FFT spectra obtained \via 5, 10, and 100 samples can be used to reconstruct the original functions.
Rows 2 to 4 in Fig.~\ref{fig:motivating-plots} plot the reconstructed functions using 5, 10, and 100 samples, respectively.
This is a visual demonstration of how well the original function is approximated using only 0.05\%, 0.1\%, and 1\% of the images in the test dataset.
In fact, had we performed the same process for all outputs of the model, the reconstructed model using 5, 10, and 100 samples would achieve test accuracies of 0.8953, 0.9877, and 0.9931, respectively, which are quite close to the original model.

Our goal here is not to reconstruct the original functions, rather we want \dms to directly use the calculated spectra to compare different mutants of a DNN.
This is possible thanks to the fact that FFT spectra uniquely characterize functions and similar functions are expected to have similar amplitudes in each of the matching frequency buckets.
The last row in Fig.~\ref{fig:motivating-plots} depicts the bar charts for the first 20 frequency buckets of the FFT spectra for the outputs 2 and 5 of the model which are obtained using 100 samples.
In these diagrams, the position of each bar represents a frequency bucket, while the height of the bars represents the amplitude for that frequency.
Given that using only 100 samples the two functions are perceived as phase-shifted ``pulse functions,'' the FFT spectra of the two functions are rather similar to each other.

\dms views FFT spectra of the mutants of a DNN model as points in a multi-dimensional space, and leverages the distance between the points to calculate similarity values that are then used to cluster the mutants into groups of mutants with likely the same outcome, \eg, all survived or all killed.

\section{\dms Approach}\label{sec:approach}
\dms reduces mutation testing costs by clustering mutants based on their approximated behavior and testing a representative from each cluster instead of all mutants.
A DNN classifier and its mutants comprise functions with common inputs, \ie, one function per output.
For example, a 10-class classifier DNN model, or any of its mutants, is a collection of 10 functions (one output per class) with the same input.
The key insight of this paper is that DNN outputs are real-valued functions suitable for FFT analysis, enabling the computation of comparable signatures for mutant behavior.
\dms clusters mutants using the magnitude of FFTs, grouping those with similar behavior.
Specifically, it employs a similarity measure based on the maximum Euclidean distance between the FFT magnitudes of mutant outputs. This method, commonly used in digital signal processing~\cite{bib:mitrovic2010features,bib:kinnunen2010an}, must be applied carefully due to differences between sound signals and mutant outputs.
Mutants arise from small variations targeting specific neurons in a DNN.
Since individual neurons contribute to specific outputs, meaning mutations affect only a subset of outputs while others remain nearly unchanged.
Consequently, their FFTs share many harmonics, leading to zero, or extremely small, Euclidean distances.
To measure mutant distances, \dms takes the maximum Euclidean distance of the outputs, emphasizing the most impacted output.
Formally, given two mutants $\mu_1$ and $\mu_2$ with $q$ outputs and a test set $S$, the distance is defined to be {\small $\delta_S(\mu_1,\mu_2)=\max_{0\leq i< q}\left\{\sqrt{\sum_{j=0}^{|S|}(\|FFT_S(\mu_1[i])\|[j] - \|FFT_S(\mu_2[i])\|[j])^2}\right\}$}, where $FFT_S$ computes the FFT of the $i$-th output using $S$, $[.]$ denotes vector/array indexing, and $\|.\|$ represents FFT magnitude.
The Euclidean distance between FFT magnitudes is computed for each output, and their maximum is used as the mutant distance.

The sample set $S$ is randomly drawn from a \textit{tiny} subset of the test dataset, \eg, 0.1\% of data points.
For a meaningful approximation of mutant behavior, the sample should ideally include at least one instance per class.
For example, in a 10-class dataset like MNIST with 10,000 test points, each class has hundreds of instances.
A 50-point sample (\ie, 0.5\% of the dataset) should thus contain 5 points per class.
Analyzing \dms in pathological cases where certain classes are missing is left for future work.


\dms leverages FFT-based distances to cluster mutants with likely identical outcomes, \eg, all surviving or getting killed.
However, most clustering algorithms require a similarity measure rather than raw distances.
To address this, we map distance values to similarity values \via an exponentiation function, and define the similarity of mutants $\mu_1$ and $\mu_2$ as:


\begin{equation}\label{eq:similarity}
    \sigma_S(\mu_1,\mu_2)=e^{-\delta_S(\mu_1,\mu_2)}
\end{equation}

Intuitively, for a given sample set $S$, $\sigma_S(\mu_1,\mu_2)=1$, if the mutants $\mu_1,\mu_2$ are identical in behavior, and the value of the $\sigma_S$ function quickly approaches to zero as the distance between the mutants behavior grows.
This rate of shrinking in exponentiation function helps spreading similarity values between 0 and 1, better than other alternatives, \eg, reciprocal, which helps interpretability of the similarity values.

With this definition of mutant similarity, we now examine DNN mutation analysis workflows incorporating \dms as an accelerator.
Fig.~\ref{fig:dms-workflow} illustrates such a workflow.
Mutants generated by a host mutation analysis system, along with the test dataset, are passed to \dms, which returns results indicating which mutants are killed or survive.
The host system then utilizes these results for tasks like computing mutation scores or fault localization.
The following sections detail each process and artifact in the order they appear in the workflow.

\subsection{Mutation Generator}\label{sec:approach:mg}
\begin{figure}[t]
    \centering
    \resizebox{\columnwidth}{!}{
    \includegraphics{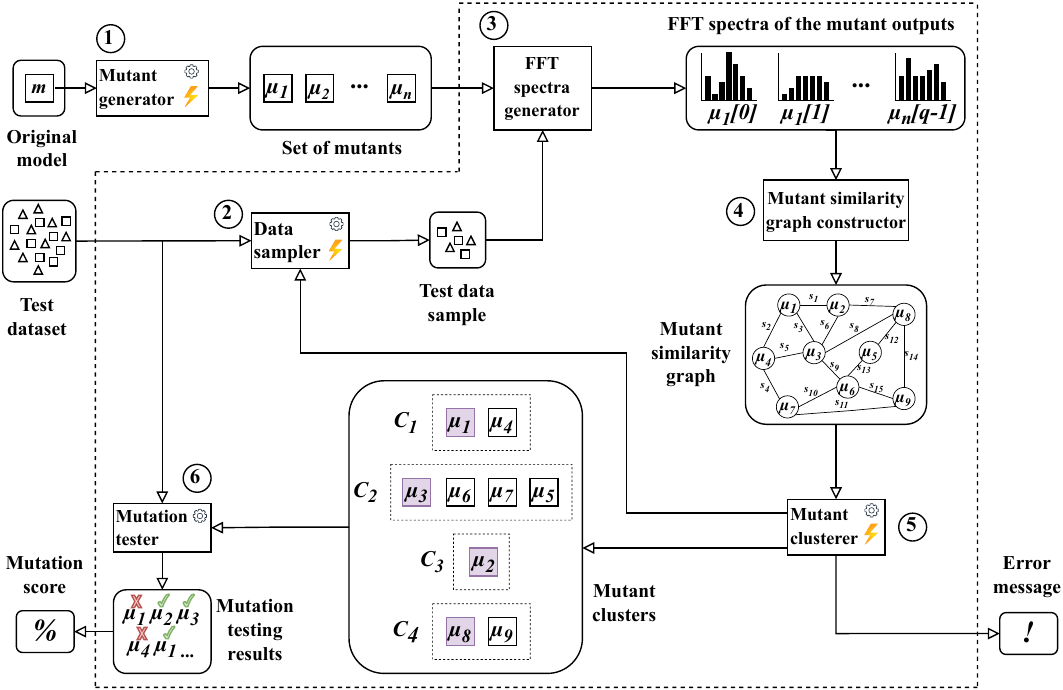}
    }
    \caption{A DNN mutation analysis workflow involving \dms. Sharp-edged rectangles denote processes; rounded ones denote data/artifacts. Arrows denote control flow. Gear and lightning bolt icons are used to annotate user-configurable and non-deterministic processes, resp. Processes and artifacts inside the area marked by dashed line are part of \dms.}
    \label{fig:dms-workflow}
\end{figure}
The mutation generator (\textcircled{1} in Fig.~\ref{fig:dms-workflow}) is a component of DNN mutation analysis systems like DeepMutation~\cite{bib:ma2018deepmutation} or DeepCrime~\cite{bib:humbatova2021deepcrime}.
It generates mutants by applying mutators to the original model or its defining code/data.
For model-level mutation, mutators directly produce mutants, while source-level mutation requires training mutated code/data to generate executable mutants.
While this paper focuses on model-level mutation, \dms is applicable to source-level analysis as well.

The mutation generator is external to \dms, providing one of its inputs.
In this paper, we employ DeepMutation's model-level mutation to generate mutants from trained models, and we use a subset of mutators, namely Gaussian Fuzzing, Weight Shuffle, Neuron Effect Block, Neuron Activation Inverse, and Neuron Switch.
These mutators, unlike DeepMutation's layer-level mutators, produce higher quality, non-trivial mutants~\cite{bib:jahangirova2020empirical} that are used in later works~\cite{bib:hu2019deepmutation++,bib:wang2019adversarial,bib:ghanbari2024decomposition}.
Mutators are parameterized and involve randomness.
In this paper, we use DeepMutation's default parameter values.

\subsection{Data Sampler}\label{sec:approach:ds}
The data sampler (\textcircled{2} in Fig.~\ref{fig:dms-workflow}) is a \dms component that selects a small subset of the test dataset for downstream FFT analysis (see~\cref{sec:approach:fft}).
Typically, only a tiny fraction, \eg, 0.1\%, of the original test dataset, which may contain thousands of data points, is sampled.
Sampling is crucial for acceleration in \dms, as it enables estimating mutant behavior without applying them to the entire test dataset.

We design \dms's data sampler to randomly select at least one instance from each class, ensuring that the FFT spectra accurately represent the mutant outputs.
The selection of an instance from a given class is currently done randomly, but various selection strategies deserve a deeper empirical analysis in future extensions of this work.
A parameter, $x$, called the \textit{data sampling rate}, determines how many instances per class should be selected.
We observed that the in-class variations of the sampled data points due to randomness, say which image of 5 in the MNIST dataset should be selected, rarely impact the outcome of the mutation analysis.
However, if too few samples are selected and a mutant misclassifies only those specific instances, the resulting FFT spectra may not accurately reflect its behavior.
Conversely, a mutant could correctly classify just the sampled instances, making the spectra appear as if from a perfect classifier.
To mitigate this randomness, we repeat all our experiments five times.

By default, \dms does not require the user to specify the $x$ value.
Instead, it iteratively invokes this component along with the mutant clusterer (see~\cref{sec:approach:cluster}) using different $x$ values from the set $\{1, 3, 5, 10, 20, 30, 40, 50, 100, 200, 300\}$, selecting 1, 3, 5, \etc., samples per class—until the user-specified mutant reduction goal is met.
However, users can disable this search loop and manually specify $x$ if they have better insights into the appropriate number of samples per class for their specific application.
For instance, in RQ1 in~\cref{sec:ans:rq1}, we disabled the loop to supply $x$ values from a predefined range and study the behavior of \dms at each value.


\subsection{FFT Spectra Generator}\label{sec:approach:fft}
Given a set of mutants $M$ and a set of test samples $S$, the FFT spectra generator (\textcircled{3} in Fig.~\ref{fig:dms-workflow}) produces FFT spectra for each output of every mutant in $M$.
These spectra are then used in the next step (see~\cref{sec:approach:gc}).
For illustration, the FFT spectra are represented as histograms labeled $\mu_1[0], \mu_1[1], \dots, \mu_n[q-1]$ in the figure.
Specifically, the histogram labeled $\mu_j[i]$ represents the FFT spectrum of output index $i$ of the mutant $\mu_j \in M$, computed using the test dataset sample $S$.
The FFT process takes a series of output values (representing a discrete function) and a set of samples as input, producing a $|S|$-dimensional vector where each element represents the magnitude of a frequency bucket.
The functions provided to this component correspond to the individual outputs of mutants in $M$, while the sample set $S$ is drawn from the test dataset.
Applying FFT on output index $i$ of mutant $\mu_j$ using $S$ is denoted as $FFT_S(\mu_j[i])$ in the definition of $\delta_S$.

A na\"{\i}ve application of FFT on mutant outputs would be costly, because given $n$ mutants, each with $q$ outputs, the FFT process with $S$ samples would need to be repeated $nq$ times.
Each round would involve applying each mutant to the data points in $S$, resulting in $|S|nq$ mutant applications, which can be expensive for mutants with large outputs.
\dms circumvents this cost by applying each mutant to each data point only once and reusing the outputs in subsequent iterations.
Specifically, \dms applies $n$ mutants to the data points in $S$, stores the results, and uses the memoized outputs when a specific mutant's output $i$ is needed during the FFT calculation.
This optimization reduces the cost of this step by a factor of $q$.

\subsection{Mutant Similarity Graph Constructor}\label{sec:approach:gc}
Mutant similarity graph constructor (\textcircled{4} in Fig.~\ref{fig:dms-workflow}) leverages the FFT spectra for the mutants outputs to construct a \textit{mutant similarity graph}, defined below.
\begin{definition}[Mutant Similarity Graph]\label{def:mutant:sim:graph}
    Mutant similarity graph is a complete weighted undirected simple graph with vertices $M$ and edges $E$. $M$ contains all and only the generated mutants, and for each $\mu_1,\mu_2\in M$, with $\mu_1\neq\mu_2$, there is an edge $(\{\mu_1,\mu_2\}, w)\in E$, where $w$, called the weight, is defined to be $\sigma_S(\mu_1,\mu_2)$ for a test dataset sample set $S$.
    \hfill\rule{1.2ex}{1.2ex}
\end{definition}

The mutant similarity graph is complete, \ie, it contains an edge between every distinct pair of mutants.
Since $\sigma_S$ is symmetric (\ie, $\sigma_S(\mu_1, \mu_2) = \sigma_S(\mu_2, \mu_1)$ for any non-empty $S$), there is a unique edge between each distinct pair, making the graph undirected and simple.
The mutant similarity graph constructor builds this graph by iterating over mutant pairs, calculating their similarities using the $\sigma_S$ function, and adding edges to an efficient C++-based graph data structure.
The $\sigma_S$ function relies on the $\delta_S$ function, which is defined using the $FFT_S$ function.
The values of $FFT_S$ are pre-calculated for each mutant output, making graph construction fast, despite the time complexity of $O(|M|^2q|S|)$, where $q$ is the number of mutant outputs.

\subsection{Mutant Clusterer}\label{sec:approach:cluster}
The mutant clusterer (\textcircled{5} in Fig.~\ref{fig:dms-workflow}) clusters nodes in the mutant similarity graph into similarity groups.
This component of \dms uses parallel hierarchical agglomerative clustering (ParHAC)~\cite{bib:dhulipala2022hierarchical}, a scalable and efficient version of the traditional hierarchical agglomerative clustering algorithm.
Scalability is important for \dms's performance, as mutant similarity graphs are typically huge.

Clustering in ParHAC is guided by the \textit{linkage threshold} parameter, denoted $\tau$.
Selecting appropriate $\tau$ value is crucial for the \dms's performance: a well-chosen $\tau$ can lead to significant time savings with minimal impact on mutation score, while a poor choice can make \dms slower than the vanilla approach for some models.
However, finding the optimal $\tau$ can be challenging, reducing the tool's usability.
We, therefore, automated finding $\tau$ value by implementing a binary search over the entire search space of valid linkage threshold values, \ie, the interval (0, 1).
The algorithm seeks a $\tau$ value that satisfies a user-defined constraint, called the \textit{mutant reduction constraint}, which specifies a desired reduction in mutant percentage, defined by the range $[l, h]$, where $0 \leq l \leq h \leq 1$.
Since one representative from each cluster is selected for testing (see~\cref{sec:approach:mt}), the reduction in mutants equals the number of clusters.
Thus, mutant reduction is calculated as $\frac{|M| - |C|}{|M|}$, where $|M|$ is the total number of mutants and $|C|$ is the number of mutant clusters.

Selecting the right $\tau$ value is closely tied to choosing the correct data sampling rate $x$, as the sampling rate affects how well FFT spectra approximate mutants' behavior.
For this, \dms performs a linear search over the set $\{1, 3, 5, 10, 20, 30, 40, 50, 100, 200, 300\}$ to select 1, 3, 5, \etc., samples from each class and then conducts a binary search for $\tau$ values for each chosen $x$.
Values 1, 3, and 5 represent \textit{small} sample sizes, 10, 20, 30, 30, 40, and 50 represent \textit{mid-size} sample sizes, while 100, 200, and 300 represent \textit{large} sample sizes.
Algorithm~\ref{alg:param-search} formalizes the steps \dms takes to coordinate the data sampler, FFT spectra generator, mutant similarity graph constructor, and mutant clusterer.
It takes the number of mutants $N$ (\ie, $|M|$) and the mutant reduction constraint $R$ as inputs, returning the clusters that satisfy $R$ or an error if the search fails.
The for-loop at Line 1 tries different $x$ values.
For each $x$, \dms performs data sampling, FFT analysis, and generates a mutant similarity graph, denoted $G_x$.
In lines 2-15, a binary search for $\tau$ values is performed to satisfy $R$.
The variables $\tau_{lo}$ and $\tau_{hi}$ represent the lower and upper bounds for $\tau$.
The while-loop at Line 4 checks if the midpoint falls within the practical range (0, 1).
Values below $10^{-5}$ are treated as zero, and values above 0.99999 as one, making the clustering algorithm to produce a single cluster or singleton clusters, respectively.
This defines the binary search termination condition, which, on average, terminates in 18.4 iterations, as per our experiments.
Lines 5-7 assign the midpoint to $\tau$, run the ParHAC clustering with $\tau$, and calculate the mutant reduction rate.
In Lines 8-14, if too many clusters are created, the upper bound is decreased; if too few, the lower bound is increased.
If a valid set of clusters is found, it is returned.
If all $x$ values are exhausted and no suitable $\tau$ is found, an error message is returned at Line 17.
\begin{figure}
    \centering
    \begin{minipage}{\columnwidth}
        \begin{algorithm}[H]
        \scriptsize
        \captionsetup{font=scriptsize}
        \caption{Parameter search procedure in \dms}\label{alg:param-search}
        \begin{algorithmic}[1]
        \REQUIRE Number of mutants $N$, mutant reduction goal $R = [l, h]$
        \ENSURE Set $C$ of clusters satisfying the mutant reduction constraint $R$, or an error message if $R$ is not achievable
        
        \FOR{$x \in \{1, 3, 5, 10, 20, 30, 40, 50, 100, 200, 300\}$}
            \STATE $\tau_{lo} \gets 0$
            \STATE $\tau_{hi} \gets 1$
            \WHILE{$10^{-5} \leq \left(\tau_{lo} + \frac{\tau_{hi} - \tau_{lo}}{2}\right)\leq 0.99999$}
                \STATE $\tau \gets \tau_{lo} + \frac{\tau_{hi} - \tau_{lo}}{2}$
                \STATE $C \gets \text{DoParHACClustering}(G_x, \tau)$
                \STATE $mut\_red\_rate \gets \frac{N - |C|}{N}$
                \IF{$mut\_red\_rate < l$}
                    \STATE $\tau_{hi} \gets \tau$
                \ELSIF{$mut\_red\_rate > h$}
                    \STATE $\tau_{lo} \gets \tau$
                \ELSE
                    \RETURN $C$
                \ENDIF
            \ENDWHILE
        \ENDFOR
        \RETURN ``Mutant reduction goal not satisfiable''
        \end{algorithmic}
        \end{algorithm}
    \end{minipage}
\end{figure}

Monotonicity of the mutant reduction rate with respect to the linkage threshold is essential for Algorithm~\ref{alg:param-search} to function properly.
While we do not prove this formally, our empirical analysis in RQ1 strongly supports this claim.
Additionally, this analysis led to selecting the range [0.26, 0.56] as a heuristic default for the mutant reduction constraint, which is expected to reduce mutation testing time by at least 10\% with no more than 5\% error in mutation score.
This allows \dms users to avoid specifying any parameters.

Mutant clusterer outputs a list of clusters with their representatives.
To create this list, it starts with an empty list, iterates over the clusters from Algorithm~\ref{alg:param-search}, randomly selects a representative node from each cluster, and adds the representative-cluster pair to the list.
Other selection strategies, such as choosing the node with the highest degree, are left for future work.
To account for randomness in representative selection and the ParHAC algorithm, we repeated the experiments five times.

\subsection{Mutation Tester}\label{sec:approach:mt}
The mutation tester (\textcircled{6} in Fig.~\ref{fig:dms-workflow}) produces the output of \dms.
It tests the cluster representatives identified by the Mutant Clusterer using the full, unsampled test dataset provided to \dms.
The output is a list of pairs $(\mu, \psi)$, where $\mu$ is a mutant and $\psi$ represents its status. 
Depending on \dms configuration, $\psi$ can be a Boolean indicating whether $\mu$ is killed or survived, or an integer representing the size of the $\mbox{killingLabels}$ set for $\mu$ as defined in~\cref{sec:back:mutation}.
\dms calculates $\psi$ for each representative and replicates it for all mutants in that cluster.
This approach, adapted from past clustering-based mutation analysis acceleration techniques~\cite{bib:hussain2008clustering,bib:ji2009clustering,bib:yu2019clustering}, marks all mutants in the cluster as killed if the representative is killed, and survived if the representative survives. 
Mutation tester is user-configurable, allowing users to define a Python function to calculate a mutant's status, either as killed/survived or by returning the size of the $\mbox{killingLabels}$ set.
For the experiments in this paper, the output is a list of pairs $(\mu, |\mbox{killingLabels}(\mu)|)$, which is used in the mutation score calculation as defined in~\cref{sec:back:mutation}.

\section{Experiments}\label{sec:experiments}
We investigate the following research questions (RQs).

\begin{itemize}
    \item \textbf{RQ1 (Monotonicity and Heuristic Parameter Values):}
    \begin{enumerate}
        \item How does the mutant reduction rate vary with the linkage threshold ($\tau$) across different data sampling rates ($x$), and is this relationship monotonic?
        \item What range of mutant reduction rates results in a maximum of 5\% error in mutation score and a minimum of 10\% reduction in mutation testing time?
    \end{enumerate}
    \item \textbf{RQ2 (Effectiveness):}
    \begin{enumerate}
        \item How does \dms, with default configuration, perform compared to the vanilla approach on larger models?
        \item How does \dms, with default configuration, compare to other baseline approaches?
    \end{enumerate}
    \item \textbf{RQ3 (Soundness and Predictive Accuracy):}
    \begin{enumerate}
        \item Does \dms provide meaningful behavioral clustering that goes beyond simple analysis of output classes over the selected test inputs?
        \item How accurate are the propagated test results from cluster representatives to other mutants?
    \end{enumerate}
\end{itemize}

The ParHAC algorithm used in \dms takes a linkage threshold parameter, $\tau$, which is tuned \via the binary search algorithm in~\cref{sec:approach}.
In the first part of RQ1, we provide empirical evidence supporting monotonicity of mutant reduction rate with respect to $\tau$, which is essential for the binary search algorithm to function correctly.
Analyzing 1,045 data points from four models, \textit{we found strong empirical evidence of a consistent monotonic relationship between the mutant reduction rate and $\tau$, with no significant fluctuations}.
Given this evidence, we expect that the aforementioned binary search to find $\tau$ value satisfying a mutant reduction constraint to terminate.

In the second part of RQ1, we use our measurement to determine heuristic default values for the mutant reduction rate, aiming to guide the binary search toward a rate that ensures no more than 5\% error in mutation score, yet at least a 10\% reduction in mutation testing time--thresholds deemed acceptable by prior work on mutation analysis acceleration~\cite{bib:feng2022mutation,bib:wang2023fine,bib:li2022higher,bib:shen2021boundary}.
Analyzing 1,045 data points from four models, \textit{we found that a mutant reduction constraint of $[0.26, 0.56]$ achieves over 10\% reduction with less than 5\% mutation score loss}.
We adopt this as the default heuristic for RQ2-3.

In RQ2, we first study the performance of \dms on more realistic DNN models.
The results provide empirical evidence on the effectiveness of \dms in \textit{accelerating mutation testing by 28.38\%, on average, at the average cost of only 0.72\%  error in mutation score}.
Next we compare \dms to three baseline approaches, namely random mutant selection~\cite{bib:ghanbari2023mutation}, boundary sample selection~\cite{bib:shen2021boundary}, and random sample selection.
We observed that, on average, \textit{\dms incurs 11.78, 15.16, and 114.36 times less mutation score error compared to random mutant selection, boundary sample selection technique, and random sample selection, respectively, while offering comparable speed-up in most of the cases}.

In RQ3, we study the effectiveness of two major components of \dms: FFT spectra generation and mutant clusterer.
We first address the concern that whether \dms reduces to an analysis of the histograms of the output classes over the selected set of test inputs.
By disabling FFT spectra generation component, and directly clustering the mutants based on their outputs, \textit{we observed that mutation score error rises $24.27\times$, confirming that FFT analysis is a key for precision in \dms}.
Next, we evaluate predictive performance of mutant clustering and \textit{found that \dms performs quite well in terms of several metrics}.

\subsection{Benchmark and Setup}\label{sec:exp:dataset-setup}

Table~\ref{tab:benchmark} lists the DNN models used in the experiments, where each row represents a model.
We have used six types of DNN architectures in our experiments: 4-layer FCNN, LetNet-5~\cite{bib:lecun1998gradient}, ResNet-10~\cite{bib:he2016deep}, MobileNetV2~\cite{bib:sandler2018mobilnetv2}, EfficientNetB2~\cite{bib:tan2019rethinking}, and RNN with LSTM layers.
We have use standard LetNet-5 architecture that represents the family of CNN models with no residual blocks, such as AlexNet~\cite{bib:krizhevsky2012imagenet} and VGGNet~\cite{bib:simonyan2015very}.
We have also implemented ResNet-10, MobileNetV2, and EfficientNetB2 based on the standard architectures, representing models with residual block of different complexity and layouts.
Our RNN architecture consists of an embedding layer, two LSTM layers, and an output layer with softmax activation.

\begin{table}
    \centering
    \caption{Benchmark for the experiments. Train \# and Test \# represent the number of data points in the train and test datasets, respectively, and Cls \# denotes the number of classes/labels in the test dataset.}\label{tab:benchmark}
    \resizebox{\columnwidth}{!}{
        \begin{tabular}{|c|l||c|r|r|r|r|r|}
            \hline
            \textbf{Architecture} & \multicolumn{1}{c||}{\textbf{Dataset}} & \textbf{Scope} & \multicolumn{1}{c|}{\textbf{Size}} & \multicolumn{1}{c|}{\textbf{Train \#}} & \multicolumn{1}{c|}{\textbf{Test \#}} & \multicolumn{1}{c|}{\textbf{Cls \#}} & \multicolumn{1}{c|}{\textbf{Test Acc.}} \\
            \hline
            \hline
            \multirow{4}{*}{\textbf{FCNN}} & \textbf{EMNIST} & \multirow{4}{*}{RQ1} & 45,676 & 124,800 & 20,800 & 26    & 0.8831 \\
        \cline{2-2}\cline{4-8}          & \textbf{FMNIST} &       & 44,860 & 60,000 & 10,000 & 10    & 0.8755 \\
        \cline{2-2}\cline{4-8}          & \textbf{KMNIST} &       & 44,860 & 60,000 & 10,000 & 10    & 0.8704 \\
        \cline{2-2}\cline{4-8}          & \textbf{MNIST} &       & 44,860 & 60,000 & 10,000 & 10    & 0.9754 \\
            \hline
            \multirow{2}{*}{\textbf{LeNet-5}} & \textbf{EMNIST} & \multirow{10}{*}{RQ2-3} & 45,786 & 124,800 & 20,800 & 26    & 0.9228 \\
        \cline{2-2}\cline{4-8}          & \textbf{SVHN} &       & 62,006 & 73,257 & 26,032 & 10    & 0.8537 \\
        \cline{1-2}\cline{4-8}    \multirow{2}{*}{\textbf{ResNet-10}} & \textbf{Caltech-101} &       & 5,033,446 & 7,316 & 1,828 & 102   & 0.721 \\
        \cline{2-2}\cline{4-8}          & \textbf{CIFAR-10} &       & 4,986,250 & 50,000 & 10,000 & 10    & 0.8236 \\
        \cline{1-2}\cline{4-8}    \multirow{2}{*}{\textbf{MobileNetV2}} & \textbf{Caltech-101} &       & 334,886 & 7,316 & 1,828 & 102   & 0.6351 \\
        \cline{2-2}\cline{4-8}          & \textbf{CIFAR-10} &       & 287,690 & 50,000 & 10,000 & 10    & 0.7185 \\
        \cline{1-2}\cline{4-8}    \multirow{2}{*}{\textbf{EfficientNetB2}} & \textbf{CIFAR-100} &       & 7,915,101 & 50,000 & 10,000 & 100    & 0.8175 \\
        \cline{2-2}\cline{4-8}          & \textbf{Dogs} &       & 7,943,281 & 12,000 & 8,580 & 120   & 0.8119 \\
        \cline{1-2}\cline{4-8}    \multirow{2}{*}{\textbf{RNN}} & \textbf{IMDB} &       & 2,642,562 & 25,000 & 25,000 & 2     & 0.812 \\
        \cline{2-2}\cline{4-8}          & \textbf{Reuters} &       & 2,648,282 & 8,982 & 2,246 & 90    & 0.642 \\
            \hline
        \end{tabular}
    }
\end{table}

We have trained these model architecture on various datasets of different complexities, such as MNIST~\cite{bib:deng2012the}, a dataset of hand-written digits with classes 0-9, Fashion MNIST~\cite{bib:xiao2017fashion} (FMNIST), a dataset of fashion items with 0-9, the digit section of Kuzushiji MNIST~\cite{bib:clanuwat2018deep} (KMNIST), a dataset of hand-written Japanese digits 0-9, and SVHN~\cite{bib:netzer2011reading}, a dataset of real-world images for 10-class classification of digits. 
We trained ResNet-10 and MobileNetV2 model architectures on more complex datasets such as CIFAR-10~\cite{bib:krizhevsky2009learning}, consisting of $32\times32$ color images in 10 different classes, and Caltech-101~\cite{bib:lazebnik2006beyond}, comprised of $224\times224$ color images belonging to 102 different object categories.
Meanwhile, we trained EfficientNetB2 on even more complex datasets CIFAR-100, consisting of $32\times32$ color images in 100 different classes, and Stanford Dogs Dataset~\cite{bib:khosha2011novel}, a 120-class subset of ImageNet~\cite{bib:deng2009imagenet}.
Lastly, we used Reuters~\cite{bib:lewis1997reuters} and IMDB~\cite{bib:maas2011learning} datasets for training the RNN models.
Reuters is a dataset for 90-class classifiers for documents with news articles, and IMDB is a dataset for binary sentiment classification for movie reviews.

We have used simpler FCCN models in RQ1, making it feasible to run \dms and vanilla mutation analysis thousands of times, while other more complex models are used in RQ2-3, as indicated in the column ``Scope'' in the table.
In the rest of the paper, we use the combination of model name and training dataset identifier to uniquely identify each of the 12 models, \eg, FCNN-FMNIST, denotes the model with FCNN architecture trained on FMNIST dataset.

We have used two identical Dell Precision workstations with AMD Ryzen Threadripper @ 2.7 GHz CPU, 1 TB of RAM, and two NVIDIA RTX A6000 GPUs to conduct our experiments.
Both machines run Ubuntu 22.04.4 LTS.

\subsection{Measures}\label{sec:exp:measures}
In vanilla mutation testing, \ie, exhaustive testing of generated mutants, \textit{mutation testing time} refers to the total time required for testing the generated mutants.
For \dms, this includes additional overhead from data sampling, FFT analysis, clustering, and parameter search.
Mutation testing time is measured in seconds but can vary by setup, so we accompany time measurements by the \textit{number of tested mutants} as well.

\textit{Speed-up}, \ie, the amount of acceleration, measures the percentage decrease in mutation testing time when using \dms, or any other mutation analysis acceleration technique, instead of the vanilla approach.
It is calculated as $\frac{T_V - T_0}{T_V}$, where $T_V$ and $T_0$ are the mutation testing times for the vanilla approach and mutation analysis acceleration technique, \eg, \dms or BSS, respectively.
A related measure, \textit{mutant reduction}, quantifies the reduction in tested mutants.
The vanilla approach always has 0 mutant reduction, as it tests all generated mutants, whereas \dms tests only one representative per cluster.
Mutant reduction is calculated by $\frac{|M| - N_0}{|M|}$, where $|M|$ is the number of generated mutants and $N_0$ is the number tested mutants.

Another measurement conducted in this paper is the \textit{mutation score} that is calculated using the formula in~\cref{sec:back:mutation}.
\textit{Mutation score error} or \textit{loss}, \ie, the percentage of deviation of accelerated mutation score from vanilla mutation score, is calculated as $\frac{|MS_V - MS_0|}{MS_V}$, where $MS_V$ is the mutation score obtained using vanilla approach, while $MS_0$ represents the mutation score obtained using an acceleration technique.

Lastly, we evaluate the predictive performance of the mutant clustering algorithm \via mean absolute error (MAE), relative MAE (RMAE), precision, recall, F1, and Matthews correlation coefficient (MCC).
MAE and RMAE are computed from the actual \vs predicted size of the \textit{killingLabels} set in the mutation score formula of~\cref{sec:back:mutation}, while the other metrics follow their standard definitions based on the confusion matrix, constructed as follows.
A mutant predicted as killed and actually killed (resp., survived) is a true (resp., false) positive.
A mutant predicted as survived and actually survived (resp., killed) is a true (resp., false) negative.
A mutant is considered \textit{survived} if its output matches that of the original model on all test data points; otherwise, it is considered \textit{killed}~\cite{bib:ma2018deepmutation}.

\subsection{Baseline Approaches}\label{sec:exp:baselines}
Random mutant selection~\cite{bib:ghanbari2023mutation}, mutation operator selection~\cite{bib:feng2022mutation,bib:wang2023fine}, test data selection~\cite{bib:shen2021boundary}, higher-order mutation~\cite{bib:li2022higher,bib:li2021second}, neuron clustering~\cite{bib:ghanbari2024decomposition}, and mutant clustering~\cite{bib:lyons2025on} are various approaches in the literature for accelerating mutation analysis.
Among them, random mutant selection (RMS), test data selection, and neuron clustering are more relevant to \dms, as they aim to reduce the cost of testing generated mutants rather than limiting their number.
Since no working implementations of these techniques were available for our setting, we reimplemented them.
We implemented RMS and boundary sample selection~\cite{bib:shen2021boundary} (BSS).
Since Ghanbari \etal~\cite{bib:ghanbari2023mutation} studied random mutant selection in mutation-based fault localization, no direct comparison of our implementation was possible, but we successfully reproduced the experiments from the original BSS paper~\cite{bib:shen2021boundary}.
This strengthened our confidence in the accuracy of our implementation.
Lyons and Ghanbari~\cite{bib:lyons2025on} have studied mutant clustering approach for single-point mutants, we observed that when applying this method on mutants impacting more than one neuron, the method becomes slower than vanilla approach due to the increased dimensionality of points to be clustered in first-order \vs higher-order mutation.
Therefore, we excluded this approach from our study.

We further compare \dms to two other baseline approaches: (1) clustering without FFT; (2) random sample selection (RSS).
The first approach performs data sampling like \dms, but it carries out clustering directly based on mutant outputs rather than their FFTs.
In the rest of this paper, we use \noFFT to denote this no-FFT variant of \dms.
RSS selects subset of test data points randomly while testing all the mutants; it selects the same number of data points that \dms does for FFT analysis.


\subsection{Results}\label{sec:exp:results}

\subsubsection{Answering RQ1}\label{sec:ans:rq1}
To address both parts of RQ1, we applied \dms to four FCNN models trained on EMNIST, FMNIST, KMNIST, and MNIST while varying the data sampling rate, $x$, and linkage threshold, $\tau$.
Specifically, we ranged $x$ over $\{1, 3, 5, 10, 20, 30, 40, 50, 100, 200, 300\}$, selecting the corresponding number of samples per class, and $\tau$ over $\{0.05, 0.1, 0.15, \dots, 0.95\}$ to systematically analyze mutant reduction, mutation score error, and timing behavior.
The first part of this RQ aims to empirically show that, across different $x$ values, the mutant reduction rate remains monotonic or constant with respect to $\tau$.
A stable reduction rate is crucial for the binary search algorithm in \dms to return correct results.
To evaluate this, we recorded the mutant reduction rate for all $(x, \tau)$ combinations, repeating the process five times to account for randomness in sampling and ParHAC clustering.
This resulted in 1,045 (=$11\times 19\times 5$) measurements per model.

\begin{wrapfigure}{l}{0.6\columnwidth}
    \centering
    \resizebox{0.6\columnwidth}{!}{
        \includegraphics{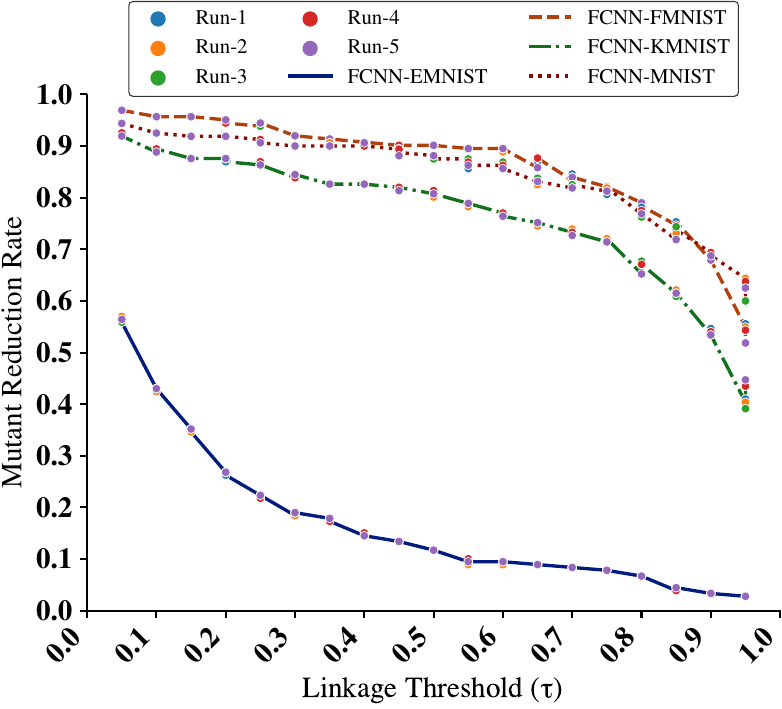}
    }
    \caption{Mutant reduction rate \vs linkage threshold when $x=1$}
    \label{fig:monotonicity}
\end{wrapfigure}

Fig.~\ref{fig:monotonicity} plots mutation reduction rate values for different $\tau$ values, when $x=1$, for four FCNN models.
Data points from 5 runs are highlighted with 5 different colors to depict the spread of mutation reduction rate for each $\tau$ value in the runs.
We can visually confirm that the mutant reduction rate is either monotonically decreasing or constant as $\tau$ increases: we observe no reversals or significant outliers across any of the runs.
The same is true for plots for all other $x$ values, which can be accessed from our replication package~\cite{bib:replica}.
To verify this visual analysis, for each $x$, \ie, for every 95 data points per model, we computed Spearman's rank correlation coefficient~\cite{bib:spearman2010the}, $\rho$, to assess the monotonicity of the relationship between $\tau$ and mutant reduction rate.
As a non-parametric measure, $\rho$ does not assume normality or linearity: $\rho = -1$ indicates a perfectly decreasing monotonic relationship, $\rho = 1$ indicates a perfectly increasing one, and $\rho = 0$ suggests no monotonic trend (\eg, fluctuation).
$\rho$ is undefined if the relationship is constant.
\begin{table}
  \centering
  \caption{Spearman's $\rho$ values for different FCNN models obtained for different sampling rates, \ie, $x$, and linkage threshold, \ie, $\tau$, values across 5 runs}\label{tab:monotonicity}
  \resizebox{\columnwidth}{!}{
    \begin{tabular}{|l|r|r|r|r|r|r|r|r|r|r|r|}
\cline{2-12}    \multicolumn{1}{r|}{} & \multicolumn{11}{c|}{\textbf{Sampling rate (samples per class)}} \\
\cline{2-12}    \multicolumn{1}{r|}{} & \multicolumn{1}{c|}{\textbf{1}} & \multicolumn{1}{c|}{\textbf{3}} & \multicolumn{1}{c|}{\textbf{5}} & \multicolumn{1}{c|}{\textbf{10}} & \multicolumn{1}{c|}{\textbf{20}} & \multicolumn{1}{c|}{\textbf{30}} & \multicolumn{1}{c|}{\textbf{40}} & \multicolumn{1}{c|}{\textbf{50}} & \multicolumn{1}{c|}{\textbf{100}} & \textbf{200} & \textbf{300} \\
    \hhline{-===========}
    \textbf{FCNN-EMNIST} & -0.999 & -0.86 & -0.832 & \multicolumn{1}{c|}{N/A} & \multicolumn{1}{c|}{N/A} & \multicolumn{1}{c|}{N/A} & \multicolumn{1}{c|}{N/A} & \multicolumn{1}{c|}{N/A} & \multicolumn{1}{c|}{N/A} & N/A   & N/A \\
    \hline
    \textbf{FCNN-KMNIST} & -0.998 & -0.999 & -0.995 & -0.823 & \multicolumn{1}{c|}{N/A} & \multicolumn{1}{c|}{N/A} & \multicolumn{1}{c|}{N/A} & \multicolumn{1}{c|}{N/A} & \multicolumn{1}{c|}{N/A} & N/A   & N/A \\
    \hline
    \textbf{FCNN-FMNIST} & -0.997 & -0.999 & -0.998 & -0.99 & -0.982 & -0.971 & -0.855 & -0.848 & \multicolumn{1}{c|}{N/A} & N/A   & N/A \\
    \hline
    \textbf{FCNN-MNIST} & -0.994 & -0.998 & -0.998 & -0.999 & -0.999 & -0.995 & -0.989 & -0.966 & -0.958 & N/A   & N/A \\
    \hline
    \end{tabular}
    }
\end{table}

Table~\ref{tab:monotonicity} reports $\rho$ values for different FCNN models.
The magnitude of Spearman's $\rho$ is commonly interpreted as follows~\cite{bib:cohen1988statistical}: $|\rho|=1$ indicates a \textit{perfect} monotonic relationship, $0.7\leq |\rho| < 1$ a \textit{strong} one, $0.4\leq |\rho| < 0.7$ a \textit{moderate} one, $0.2\leq |\rho| < 0.4$ a \textit{weak} one, and $|\rho|<0.2$ a \textit{very weak} or \textit{non-monotonic} relationship.
As shown in the table, all calculated $\rho$ magnitudes exceed 0.7, mostly approaching 1.
The negative sign confirms that mutant reduction rate decreases monotonically with $\tau$, as higher linkage thresholds make clustering stricter, leading to more small clusters.
Undefined $\rho$ values (denoted ``N/A'') indicate a constant mutant reduction rate with respect to $\tau$.
In both cases, the results provide strong empirical evidence that mutant reduction rate remains stable as $\tau$ varies.
This supports the correctness of \dms’s binary search algorithm for selecting $\tau$ under a mutant reduction constraint.
Another observation we make is that as the data sampling rate increases, the relationship between mutant reduction rate and $\tau$ becomes constant for more models, as seen in the growing number of ``N/A'' cells from top to bottom in the table.
This occurs because higher sampling rates improve FFT's approximation of mutant behavior, revealing more differences and making clustering difficult regardless of $\tau$.
In practice, a plateau in mutant reduction rate is undesirable, as it may render binary search for a suitable $\tau$ ineffective. To mitigate this, \dms searches from lower to higher sampling rates, aiming to explore diverse search spaces before performing binary search in a space that may collapse to a single value.

\begin{wrapfigure}{l}{0.55\columnwidth}
    \resizebox{0.55\columnwidth}{!}{
        \includegraphics{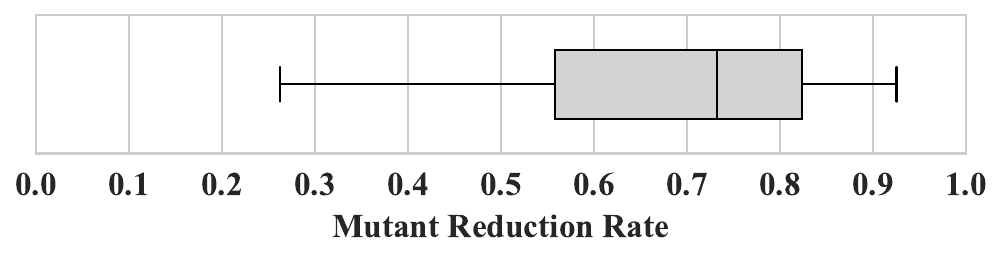}
    }
    \caption{Box-plot visualizing the mutant reduction rates that result in no more than 5\% mutation score error and at least 10\% mutation testing speed gain}\label{fig:r_val_box}
\end{wrapfigure}
An engineering goal in \dms's design is to minimize dependence on user-defined parameters, such as $x$ and $\tau$, by determining them automatically.
Finding the right mutant reduction constraint to balance mutation score error and speed-up is challenging and could impact usability.
To address this, we analyzed measurement data from this RQ: for each model and 1,045 $x$ and $\tau$ combinations across 5 runs, we recorded mutation score error, mutation testing time (including data sampling and clustering), and mutant reduction rate.
The box plot in Fig.~\ref{fig:r_val_box} highlights 124 of 4,180 values (1,045 per model for 4 models) that yield no more than 5\% mutation score loss and at least 10\% speed-up.

The minimum, 25\textsuperscript{th} percentile, median, 75\textsuperscript{th} percentile, and maximum mutant reduction rates are 0.26, 0.56, 0.73, 0.82, and 0.93, respectively.
When setting the default mutant reduction goal, any value between the minimum and maximum could be chosen.
However, just because \dms reduces approximately 82\% of mutants for one model does not mean that it will achieve the same for another without greater mutation score loss.
Notably, a mutant reduction constraint of [0.26, 0.56] yields over 10\% reduction with less than 5\% mutation score loss for more models than a higher constraint, \eg, [0.56, 0.73].
Likewise, [0.56, 0.73] would work for more models than a range containing larger values, \eg, [0.73, 0.93].
Thus, we conservatively select [0.26, 0.56], spanning the minimum to the 25\textsuperscript{th} percentile.
As we will see in RQ2, this constraint generalizes to more complex models, maintaining similar trends in mutation score loss and speed gain (see~\cref{sec:ans:rq2}).
Prior mutation analysis acceleration~\cite{bib:feng2022mutation,bib:wang2023fine,bib:li2022higher,bib:shen2021boundary} treats $\leq$5\% error as acceptable, because mutation score is itself a proxy for test adequacy and small deviations rarely affect downstream adequacy decisions.
Errors below 1\%, as consistently observed with \dms, can thus be considered negligible in practice.

A second view of this data wherein we plot plot mutation score error \vs mutant reduction rate is accessible from~\cite{bib:replica}.
This corroborates that the majority of data points corresponding to $\leq$5\% mutation score error fall in the range [0.26, 0.56].
To further validate these findings, we conducted another Spearman's analysis between the mutation score error and MAE/RMAE.
We found strong, statistically significant positive correlations, confirming that global deviations in mutation scores mirror fine-grained deviations at the mutant level.
Because mutation score error is strongly, positively correlated with both MAE and RMAE, the region [0.26, 0.56] can also be expected to exhibit the low MAE/RMAE, while achieving at least 10\% speed-up.
See~\cite{bib:replica} for more details.

\subsubsection{Answering RQ2}\label{sec:ans:rq2}
\begin{table}[h]
    \centering
    \caption{\dms (with default configuration) results on larger models}\label{tab:larger-models}
    \resizebox{\columnwidth}{!}{
        \begin{tabular}{|r|l|r|r|r|r|r|r|r|r|r|}
    \hline
    \multicolumn{2}{|c|}{\textbf{Model Under Test}} & \multicolumn{3}{c|}{\textbf{Mutation Score Error}} & \multicolumn{3}{c|}{\textbf{Mutants Reduced}} & \multicolumn{3}{c|}{\textbf{Total Time Reduction}} \\
    \hline
    \multicolumn{1}{|c|}{\textbf{Architecture}} & \multicolumn{1}{c|}{\textbf{Dataset}} & \multicolumn{1}{c|}{\textbf{Avg}} & \multicolumn{1}{c|}{\textbf{Min}} & \multicolumn{1}{c|}{\textbf{Max}} & \multicolumn{1}{c|}{\textbf{Avg}} & \multicolumn{1}{c|}{\textbf{Min}} & \multicolumn{1}{c|}{\textbf{Max}} & \multicolumn{1}{c|}{\textbf{Avg}} & \multicolumn{1}{c|}{\textbf{Min}} & \multicolumn{1}{c|}{\textbf{Max}} \\
    \hline
    \hline
    \multicolumn{1}{|c|}{\multirow{2}{*}{\textbf{LeNet-5}}} & \textbf{EMNIST} & 0.17\% & 0.07\% & 0.29\% & 37.23\% & 36.76\% & 37.55\% & 35.58\% & 32.96\% & 37.40\% \\
\cline{2-11}          & \textbf{SVHN} & 0.71\% & 0.32\% & 1.05\% & 49.28\% & 47.68\% & 51.48\% & 29.34\% & 27.07\% & 31.35\% \\
    \hline
    \multicolumn{1}{|c|}{\multirow{2}{*}{\textbf{ResNet-10}}} & \textbf{Caltech-101} & 0.81\% & 0.29\% & 1.39\% & 26.74\% & 25.04\% & 29.41\% & 20.03\% & 15.22\% & 23.36\% \\
\cline{2-11}          & \textbf{CIFAR-10} & 0.08\% & 0.01\% & 0.18\% & 42.47\% & 41.34\% & 44.06\% & 38.29\% & 35.36\% & 41.11\% \\
    \hline
    \multicolumn{1}{|c|}{\multirow{2}{*}{\textbf{MobileNetV2}}} & \textbf{Caltech-101} & 0.40\% & 0.18\% & 0.55\% & 28.07\% & 27.94\% & 28.19\% & 12.05\% & 10.07\% & 13.94\% \\
\cline{2-11}          & \textbf{CIFAR-10} & 0.00\% & 0.00\% & 0.00\% & 36.38\% & 36.33\% & 36.41\% & 22.87\% & 20.69\% & 26.56\% \\
    \hline
    \multicolumn{1}{|c|}{\multirow{2}{*}{\textbf{EfficientNetB2}}} & \textbf{CIFAR-100} & 0.60\% & 0.08\% & 1.18\% & 27.33\% & 25.65\% & 29.96\% & 20.52\% & 15.75\% & 23.83\% \\
\cline{2-11}          & \textbf{Dogs} & 0.87\% & 0.35\% & 1.45\% & 26.32\% & 24.65\% & 28.93\% & 23.76\% & 19.17\% & 26.93\% \\
    \hline
    \multicolumn{1}{|c|}{\multirow{2}{*}{\textbf{RNN}}} & \textbf{IMDB} & 1.93\% & 0.00\% & 3.53\% & 47.26\% & 39.03\% & 52.82\% & 47.13\% & 35.65\% & 49.20\% \\
\cline{2-11}          & \textbf{Reuters} & 1.65\% & 0.90\% & 3.00\% & 35.97\% & 28.93\% & 54.05\% & 34.25\% & 27.39\% & 51.23\% \\
    \hline
    \end{tabular}
        
    }
\end{table}

To evaluate \dms's effectiveness, with its default configuration, on several larger models, we enabled automatic linkage threshold search of the tool.
Table~\ref{tab:larger-models} presents our results.
The first two columns list model architecture and dataset, while subsequent columns report: (1) ``Mutation Score Error'' (average, minimum, and maximum in percent), (2) ``Reduced Mutants'' (average, minimum, and maximum in percent), and (3) ``Total Time Reduction'' (average, minimum, and maximum, including clustering, linkage threshold search, and sampling rate search). 
Results summarize five runs per model.
Overall, \dms reduces the number of tested mutants by 35.71\%, on average, which translates to 28.38\% average reduction in mutation testing time, at the cost of only 0.72\% error in average mutation score.
\dms achieves this by sampling only one data point per class for all but one, \ie, RNN-IMDB, model.
For this model, the linear sampling search ends up selecting 10 samples per class in order to satisfy the default mutant reduction constraint [0.26, 0.56].
This is because IDBM has only two classes, and with few data points, \ie, 1, 3, or 5 samples per class the mutant reduction rate goes above 56\%, prompting \dms to increase data sampling rate to increase its accuracy so that it can better distinguish mutants from one another and avoid clustering less non-equivalent mutants into one group.
Despite this, \dms is able to reduce the total mutation testing time (which includes the time needed to try smaller sampling rates) by at least 35.65\%.

The average number of binary search tries for finding the linkage threshold has been 18.38 (min: 1, max: 104).
The outlier number of tries, \ie, 104, belongs to RNN-IMDB, which requires three rounds of full binary search before finding the right linkage threshold at a sampling rate of 10 samples per class.
In addition, given that the number of FFT frequency buckets used for clustering is equal to the number of test data points selected to compute the spectra, and the number of classes reported in Table~\ref{tab:benchmark}, the number of frequency buckets used in our experiments ranged from 10 (for 10-class classifiers with $x=1$) to 120 (for the 120-class classifier with $x=1$).



\begin{table}[h]
  \centering
  \caption{Comparison of \dms to the baseline approaches}\label{tab:dms-vs-sota}
  \resizebox{\columnwidth}{!}{
        \begin{tabular}{|c|l||r|r|r|r|r|r|r|r|}
    \hline
    \multicolumn{2}{|c||}{\textbf{Model Under Test}} & \multicolumn{4}{c|}{\textbf{Average Mutation Score Error}} & \multicolumn{4}{c|}{\textbf{Average Time Reduction}} \\
    \hline
    \multicolumn{1}{|c|}{\textbf{Architecture}} & \multicolumn{1}{c||}{\textbf{Dataset}} & \multicolumn{1}{c|}{\textbf{RMS}} & \multicolumn{1}{c|}{\textbf{RSS}} & \multicolumn{1}{c|}{\textbf{BSS}} & \multicolumn{1}{c|}{\textbf{\dms}} & \multicolumn{1}{c|}{\textbf{RMS}} & \multicolumn{1}{c|}{\textbf{RSS}} & \multicolumn{1}{c|}{\textbf{BSS}} & \multicolumn{1}{c|}{\textbf{\dms}} \\
    \hline
    \hline
    \multicolumn{1}{|c|}{\multirow{2}{*}{\textbf{LeNet-5}}} & \textbf{EMNIST} & 3.77\% & 96.48\% & 7.20\% & 0.17\% & 32.69\% & 69.31\% & 60.54\% & 35.58\% \\
\cline{2-10}          & \textbf{SVHN}  & 4.08\% & 97.98\% & 1.46\% & 0.71\% & 30.35\% & 73.65\% & 58.72\% & 29.34\% \\
    \hline
    \multicolumn{1}{|c|}{\multirow{2}{*}{\textbf{ResNet-10}}} & \textbf{Caltech-101} & 3.35\% & 84.76\% & 10.83\% & 0.81\% & 26.97\% & 73.81\% & 73.51\% & 20.03\% \\
\cline{2-10}          & \textbf{CIFAR-10} & 2.20\% & 99.63\% & 0.33\% & 0.08\% & 27.91\% & 64.68\% & 68.05\% & 38.29\% \\
    \hline
    \multicolumn{1}{|c|}{\multirow{2}{*}{\textbf{MobileNetV2}}} & \textbf{Caltech-101} & 3.45\% & 93.32\% & 11.45\% & 0.40\% & 22.09\% & 25.33\% & 50.07\% & 12.05\% \\
\cline{2-10}          & \textbf{CIFAR-10} & 3.29\% & 88\%  & 0\%   & 0\%   & 31.81\% & 46.80\% & 39.70\% & 22.87\% \\
    \hline
    \multicolumn{1}{|c|}{\multirow{2}{*}{\textbf{EfficientNetB2}}} & \textbf{Dogs}  & 3.94\% & 53.26\% & 9.68\% & 0.60\% & 30.18\% & 52.38\% & 79.10\% & 20.52\% \\
\cline{2-10}          & \textbf{CIFAR-100} & 1.36\% & 65.43\% & 11.50\% & 0.87\% & 22.92\% & 56.52\% & 70.48\% & 23.76\% \\
    \hline
    \multicolumn{1}{|c|}{\multirow{2}{*}{\textbf{RNN}}} & \textbf{IMDB}  & 32.42\% & 63.65\% & 32.47\% & 1.93\% & 68.57\% & 97.42\% & 96.01\% & 47.13\% \\
\cline{2-10}          & \textbf{Reuters} & 27.67\% & 83.16\% & 24.46\% & 1.65\% & 69.20\% & 76.44\% & 78.36\% & 34.25\% \\
    \hline
    \end{tabular}
    }
\end{table}%

Table~\ref{tab:dms-vs-sota} reports average mutation score error and average time reduction for four approaches RMS, RSS, BSS, and \dms (with its default configuration).
For RMS, we selected 75\% of the generated mutants, as the previous work~\cite{bib:ghanbari2023mutation} reports less performance drop for this setting.
Similarly, for BSS, we selected the default threshold value of 10, which is reported to perform best according to the past work~\cite{bib:shen2021boundary}.
Meanwhile, for RSS, we selected one sample from each class, except for RNN-IMDB, where we sampled 10 data points from each class.
Each tool has been executed five times and the table reports average values.
The readers are referred to the replication package~\cite{bib:replica} for more details about the values, as well as RMS performance for other selection percentages.

As we can see, in this dataset, RSS consistently yields highest speed-up (63.64\%, on average), as it significantly reduces the size of the test data, which is also why we get the highest loss (82.57\%, on average).
BSS sits in the second place, reducing total mutation testing time by 67.45\%, on average.
But this gain is at the cost of average 10.94\% loss.
Surprisingly, RMS resulted in a smaller loss of 8.50\%, on average.
This observation calls for further study on the power of RMS approach.
We can also see that \dms consistently outperforms other techniques in terms of average loss (averaging 0.72\%), and its average speed-up is comparable to that of RMS (\ie, 36.27\%).

Lastly, we would like to emphasize that different techniques offer distinct strengths and weaknesses, leading to a speed \vs accuracy tradeoff.
When speed is prioritized over accuracy, RMS or BSS are well-suited, particularly for tasks such as test data augmentation with iterative methods like genetic algorithms.
In such cases, the mutation score from RMS or BSS can serve as a fitness function to guide dataset improvement across iterations, after which test suite quality can be evaluated more accurately using methods like \dms or vanilla.

\subsubsection{Answering RQ3}\label{sec:ans:rq3}
For the first part of this RQ, we disabled \dms's FFT spectra generation component to do clustering directly based on mutant outputs.
Table~\ref{tab:noFFT} reports our results in terms of average loss in mutation score, mutant reduction, and speed-up.
\begin{table}[h]
  \centering
  \caption{Comparing \dms to its no-FFT variant, \noFFT}\label{tab:noFFT}%
  \resizebox{0.9\columnwidth}{!}{
    \begin{tabular}{|c|l||r|r|r|r|r|r|}
    \hline
    \multicolumn{2}{|c||}{\textbf{Model Under Test}} & \multicolumn{2}{c|}{\textbf{Avg. Loss}} & \multicolumn{2}{c|}{\textbf{Avg. Mut. Red.}} & \multicolumn{2}{c|}{\textbf{Avg. Speed-up}} \\
    \hline
    \textbf{Architecture} & \multicolumn{1}{c||}{\textbf{Dataset}} & \multicolumn{1}{c|}{\textbf{\dms}} & \multicolumn{1}{c|}{\textbf{\noFFT}} & \multicolumn{1}{c|}{\textbf{\dms}} & \multicolumn{1}{c|}{\textbf{\noFFT}} & \multicolumn{1}{c|}{\textbf{\dms}} & \multicolumn{1}{c|}{\textbf{\noFFT}} \\
    \hline
    \hline
    \multirow{2}{*}{\textbf{LeNet-5}} & \textbf{EMNIST} & 0.17\% & 6.09\% & 37.23\% & 30.83\% & 35.58\% & 32.15\% \\
\cline{2-8}          & \textbf{SVHN} & 0.71\% & 10.58\% & 49.28\% & 27.00\% & 29.34\% & 19.90\% \\
    \hline
    \multirow{2}{*}{\textbf{ResNet-10}} & \textbf{Caltech-101} & 0.81\% & 7.94\% & 26.74\% & 33.62\% & 20.03\% & 29.41\% \\
\cline{2-8}          & \textbf{CIFAR-10} & 0.08\% & 5.26\% & 42.47\% & 34.52\% & 38.29\% & 33.13\% \\
    \hline
    \multirow{2}{*}{\textbf{MobileNetV2}} & \textbf{Caltech-101} & 0.40\% & 6.49\% & 28.07\% & 48.69\% & 12.05\% & 22.69\% \\
\cline{2-8}          & \textbf{CIFAR-10} & 0.00\% & 5.80\% & 36.38\% & 36.37\% & 22.87\% & 21.23\% \\
    \hline
    \multirow{2}{*}{\textbf{EfficientNetB2}} & \textbf{Dogs} & 0.87\% & 26.19\% & 26.32\% & 34.55\% & 23.76\% & 33.80\% \\
\cline{2-8}          & \textbf{CIFAR-100} & 0.60\% & 32.71\% & 27.33\% & 34.70\% & 20.52\% & 31.55\% \\
    \hline
    \multirow{2}{*}{\textbf{RNN}} & \textbf{IMDB} & 1.93\% & 43.29\% & 47.26\% & 46.60\% & 47.13\% & 45.98\% \\
\cline{2-8}          & \textbf{Reuters} & 1.65\% & 30.90\% & 35.97\% & 35.04\% & 34.25\% & 33.65\% \\
    \hline
    \end{tabular}%
    }
\end{table}%

While \noFFT achieves higher reduction rates (\eg, MobileNetV2-Caltech-101, 48.69\% vs. 28.07\%), this gain comes at the expense of drastically larger mutation score error, often an order of magnitude larger than \dms.
For instance, on EfficientNetB2-CIFAR-100, \noFFT incurs 32.71\% error compared to only 0.60\% with \dms. This pattern seems to generalize: \dms maintains sub-1\% error across most models, whereas \noFFT always exceeds 5\%.
The $24.27\times$ average difference demonstrates that FFT analysis is \textit{not} a cosmetic step but the main driver of precision.
In practice, this means that simply clustering raw mutant outputs is too coarse to preserve behavioral fidelity, especially on high-class datasets.
\begin{table}[h!]
  \centering
  \caption{Predictive performance of \dms's mutant clusterer component}\label{tab:predPref}
  \resizebox{\columnwidth}{!}{
    \begin{tabular}{|c|l||r|r||r|r|r|r|r|r|r|r|}
    \hline
    \multicolumn{2}{|c||}{\textbf{Model Under Test}} & \multicolumn{1}{c|}{\multirow{2}{*}{\textbf{MAE}}} & \multicolumn{1}{c||}{\multirow{2}{*}{\textbf{RMAE}}} & \multicolumn{4}{c|}{\textbf{Confusion Matrix}} & \multicolumn{1}{c|}{\multirow{2}{*}{\textbf{Prec.}}} & \multicolumn{1}{c|}{\multirow{2}{*}{\textbf{Rec.}}} & \multicolumn{1}{c|}{\multirow{2}{*}{\textbf{F1}}} & \multicolumn{1}{c|}{\multirow{2}{*}{\textbf{MCC}}} \\
\cline{1-2}\cline{5-8}    \textbf{Architecture} & \multicolumn{1}{c||}{\textbf{Dataset}} &       &       & \multicolumn{1}{c|}{\textbf{TP}} & \multicolumn{1}{c|}{\textbf{FP}} & \multicolumn{1}{c|}{\textbf{TN}} & \multicolumn{1}{c|}{\textbf{FN}} &       &       &       &  \\
    \hline
    \hline
    \multirow{2}{*}{\textbf{LeNet-5}} & \textbf{EMNIST} & 0.731 & 0.038 & 253   & 0     & 0     & 0     & 1     & 1     & 1     & N/A \\
\cline{2-12}          & \textbf{SVHN} & 0.536 & 0.107 & 237   & 0     & 0     & 0     & 1     & 1     & 1     & N/A \\
    \hline
    \multirow{2}{*}{\textbf{ResNet-10}} & \textbf{Caltech-101} & 1.452 & 0.060 & 3111  & 0     & 0     & 0     & 1     & 1     & 1     & N/A \\
\cline{2-12}          & \textbf{CIFAR-10} & 0.114 & 0.013 & 3021  & 0     & 0     & 0     & 1     & 1     & 1     & N/A \\
    \hline
    \multirow{2}{*}{\textbf{MobileNetV2}} & \textbf{Caltech-101} & 1.308 & 0.048 & 2824  & 0     & 0     & 0     & 1     & 1     & 1     & N/A \\
\cline{2-12}          & \textbf{CIFAR-10} & 0.000 & 0.000 & 2733  & 0     & 0     & 0     & 1     & 1     & 1     & N/A \\
    \hline
    \multirow{2}{*}{\textbf{EfficientNetB2}} & \textbf{Dogs} & 1.418 & 0.059 & 6049  & 0     & 0     & 0     & 1     & 1     & 1     & 1 \\
\cline{2-12}          & \textbf{CIFAR-100} & 1.384 & 0.051 & 6035  & 0     & 0     & 0     & 1     & 1     & 1     & 1 \\
    \hline
    \multirow{2}{*}{\textbf{RNN}} & \textbf{IMDB} & 0.175 & 0.175 & 225   & 19    & 227   & 44    & 0.922 & 0.836 & 0.877 & 0.760 \\
\cline{2-12}          & \textbf{Reuters} & 0.033 & 0.033 & 140   & 3     & 459   & 3     & 0.979 & 0.979 & 0.979 & 0.973 \\
    \hline
    \end{tabular}
    }
\end{table}

\dms propagates mutation testing results of a representative to all other mutants that can introduce imprecision.
To quantify this, we measure MAE and RMAE for predicted \vs actual killingLabels set sizes (see~\cref{sec:back:mutation}), and we also computed precision, recall, F1, and MCC when treating mutation outcomes in the classical kill/survive sense (see~\ref{sec:exp:measures}).
Table~\ref{tab:predPref} presents these results for one run, wherein we use ``N/A'' to show division by zero when calculating MCC. 
When MAE and RMAE are considered, we observed that \dms's clustering algorithm consistently performs well across all benchmark models: for CIFAR-10 models, the clusterer achieves near-perfect accuracy, with virtually zero error on MobileNetV2.
Even on more complex datasets, such as Caltech-101 and CIFAR-100, the relative error remains below 6\%, which is an evidence for generalization to more complex datasets.
On LeNet-5 and RNN models, MAE and RMAE likewise remain low, which is an indication of consistent predictive performance.
When considering precision, recall, F1, and MCC, we observed that most of the mutants are killed based on classic standard, which was expected~\cite{bib:ma2018deepmutation,bib:shen2021boundary}.
In RNN models, though, we have killed and survived mutants.
In these cases, the clusterer performs quite well with a precision above 90\% and quite high recall, F1, and MCC.
Together, these results support that \dms not only preserves mutation score at the mutation score error level but also provides reliable fine-grained predictions of individual mutant outcomes.



 \subsection{Threats to Validity}\label{sec:exp:discussion}

DNN mutation analysis is subject to many random factors~\cite{bib:jahangirova2020empirical,bib:tambon2023probabilistic}, \eg, in the case of DeepMutation, the randomness in training and mutation generation might impact the mutation score.
In this paper, our goal is not to study the reliability of mutation score as a measure for test quality, rather we aim to speed up a single run of mutation testing, so we have assumed that a fixed trained model and a set of mutants generated based off of that model are given.
However, as discussed in~\cref{sec:approach} \dms itself has non-deterministic components.
To account for this randomness, we have repeated our experiments five times and reported averaged results.

The set of DNN models studied in this paper is not a complete representative of the models across all applications.
Given the limited resources and time, a set of models with varying sizes and structures, that the we believe are representative of the models studied in the DNN mutation analysis literature~\cite{bib:ma2018deepmutation,bib:hu2019deepmutation++,bib:humbatova2021deepcrime,bib:jahangirova2020empirical,bib:wang2019adversarial} are selected.
For example, the benchmark contains models representing CNN models with/without residual blocks, and it also contains RNN models.
\dms is open sourced, so the research community can apply it on different models and study its effectiveness on a broader range of models and mutation analysis applications.

We have obtained default values for the mutant reduction range, as well as the search space for data sampling rate, based on our observation of a limited number of models.
Although these default values work satisfactorily for 10 large and complex models studies in this paper, there is no guarantee that they will generalize to other models.
Alternatively, there might be different values performing better in terms of mutation score error or mutant reduction rate.
\dms receives user-defined values for mutant reduction constraint and data sampling rate, so the user can try different values in case it fails.



\section{Related Work}\label{sec:related}
Motivated by the many potential practical applications of DNN mutation analysis, researcher have proposed several methods for reducing the costs of this process.
Some of these methods are more or less directly ported from mutation analysis of conventional programs into DNNs, while others take advantage of the unique characteristics of DNNs.
For example, Feng \etal~\cite{bib:feng2022mutation} and Wang \etal~\cite{bib:wang2023fine} identify sufficient subsets of existing mutators from the literature~\cite{bib:ma2018deepmutation,bib:shen2018munn} to avoid redundant mutants.
Ghanbari \etal~\cite{bib:ghanbari2023mutation} adopts random mutant selection, and the technique presented by Li \etal~\cite{bib:li2022higher,bib:li2021second} is based on the idea of reducing the cost of mutation testing with higher-order mutants.
Meanwhile, the technique introduced by Shen \etal~\cite{bib:shen2021boundary} relies on the assumption that mutants of a DNN model are more likely to produce different results for the test data points around the decision boundary of the model, so it samples the test data that lie at the decision boundary of the model under test.
Recently, Ghanbari~\cite{bib:ghanbari2024decomposition} proposed to accelerate mutation analysis by generating fewer mutants through clustering the neurons.
This approach together with mutant clustering based on the similarity of mutated weights have also been studied recently~\cite{bib:lyons2025on}.

Klabunde \etal~\cite{bib:klabunde2023similarity} survey broader methods for measuring DNN functional and representational similarities, which could also be applied to mutant clustering, but it is unclear if these methods could operate with a very few sample data points, \eg, only 0.1\% of the size of the test dataset, as FFT-based similarity method does.
We leave investigating these alternatives in such cases for future work.

\section{Conclusions}\label{sec:conclusions}
We propose a novel technique and tool, named \dms, for accelerating DNN mutation analysis.
The idea behind \dms is that DNN outputs, being real-valued functions, are amenable to FFT analysis that can be used to calculate a comparable signature of the overall behavior of the mutants using only a few test data samples.
\dms uses this to cluster and mutant and test a representative from each cluster instead of testing all of the mutants.
\totalDNNs DNN models, of varying sizes, trained on datasets with various complexities, have been used to study \dms from different perspectives.
The results suggest that \dms reduces the number of mutants to be tested by 35.71\%, on average, which translates into 28.38\% reduction in end-to-end mutation testing time, while incurring only an average of 0.72\% mutation score error.
We further compared \dms to three baselines techniques, RMS, BSS, and RSS.
We observed that while RMS, BSS, and RSS are 1.28, 2.38, and 2.91 times, respectively, faster than \dms, \dms incurs 11.78, 15.16, and 114.36 times less mutation score error than RMS, BSS, and RSS, respectively.
Lastly, we demonstrated that \dms does not collapse to a trivial histogram-based analysis, and we confirmed that the clustering results can be propagated with high predictive accuracy.



\section*{Data Availability}
Our measurements, as well as implementations of \dms, RMS, BSS, and RSS, are available in a replication package~\cite{bib:replica}.

\section*{Acknowledgments}
We would like to thank Anonymous ASE 2025 Reviewers for their valuable feedback.
The first author is partially supported by the NSF grant \#2446393.
Any opinions, findings, and conclusions or recommendations expressed in this material are those of the authors and do not necessarily reflect the views of the NSF or their employers.

\bibliographystyle{IEEETran}
\bibliography{main}

\end{document}